\pdfoutput=1
\documentclass[preprint,aps,showpacs,preprintnumbers,amsmath,amssymb]{revtex4}
\usepackage{graphicx}
\usepackage{dcolumn}
\usepackage{bm}
\usepackage{color}
\begin{document}
\title{Non-Hermitian approach of  edge states and quantum transport\\ in 
 a magnetic field}
\author{B. Ostahie$^{1,2}$, M. Ni\c ta$^{1}$ and A. Aldea$^{1}$}
\affiliation{$^1$ National Institute of Materials Physics, 
77125 Bucharest-Magurele, Romania. \\
$^2$ Department of Physics, University of Bucharest}

\begin{abstract}
We develop a manifest non-Hermitian approach of spectral and transport 
properties of two-dimensional  mesoscopic systems in strong magnetic field. 
The finite system to which several terminals are attached  constitutes
an open system that can be described by an effective Hamiltonian.
The life time of the quantum states expressed by the energy 
imaginary part depends specifically on the lead-system
coupling and  makes the difference among  three regimes: 
resonant, integer quantum Hall effect and superradiant.  
The discussion is carried on in terms of edge state life time
in different gaps, channel formation, role of hybridization,
transmission coefficients quantization.
A toy model helps in understanding non-Hermitian aspects in open 
systems.
\end{abstract}
\pacs{73.23.-b, 05.60.Gg, 73.43.Cd}
\maketitle
\section{Introduction}
Integer quantum Hall effect (IQHE) is a topological transport effect exhibited
by the two-dimensional (2D) electron systems when subjected to a 
strong perpendicular magnetic field.
The theoretical understanding of this effect followed two distinct routes.
The first one starts with the Kubo formula for the
conductivity of an infinite system, and adapts it for the case when the  
Fermi level sits in an energy  gap between Landau levels \cite{Streda}.
The conductivity can be expressed as an integral  over the Brillouin zone
of the Berry curvature, and it turns out to be  (in units $e^2/h$)
a topological invariant known as the first Chern number specific 
to the given gap \cite{TKNN}.

The second route, which is closer to the physical realization of the 
quantum Hall devices, puts forward the role
of the {\it edge states} induced by the strong  magnetic field
in {\it confined} 2D electron systems. The equivalence 
of the two approaches is discussed in terms of bulk-edge correspondence.

The edge states exhibit chirality  due to the broken time-reversal  
symmetry of the Hamiltonian and fill the gaps between the usual 
Landau bands.
In this picture, as long as the Fermi level $E_F$ sits in a gap,
the current is carried by  edge states. 
The transverse conductance exhibits 
plateaus $G_{xy}(E_F)=Ne^2/h$, where the integer $N$ represents  
the number of edge states taking part in the transport.
This conventional description of the IQHE is supported by the 
well-known energy spectrum of the  2D electron gas in the  
presence of hard wall edges (with strip or annular geometry),
which is  sketched in Fig.1 as function of the position of the guiding 
center $X_0$ \cite{Halperin, Prange, MacD}.
The spectrum exhibits degenerate Landau-type levels in the middle
of the strip, but shows dispersion close to the edges at $X_0=\pm L/2$.
The states located near the walls (i.e., the edge states)  fill the  gaps, 
the number of edge states at the Fermi level depending obviously on 
the gap index  where $E_F$ sits.

The description provided by  the strip model is  heuristic 
and far from experimental circumstances. It becomes more realistic 
by taking two more steps:
i) to replace the infinite strip with  a finite sample, described 
by a 2D plaquette with vanishing boundary conditions all around, 
and next ii) to open the  system by attaching  four semi-infinite leads 
necessary for injecting/collecting the electron current, and also for 
measuring the Hall voltage. If the  leads  are attached to the 
margin of the sample, the edge states {\it  hybridize} with the electron modes 
in the semi-infinite leads and give rise to {\it scattering } states, which 
describe the electron propagation from one reservoir to other ones 
by passing through  the sample \cite{Baranger, Janssen}. 
\begin{figure}[htb]
\includegraphics[angle=-00,width=0.3\textwidth]{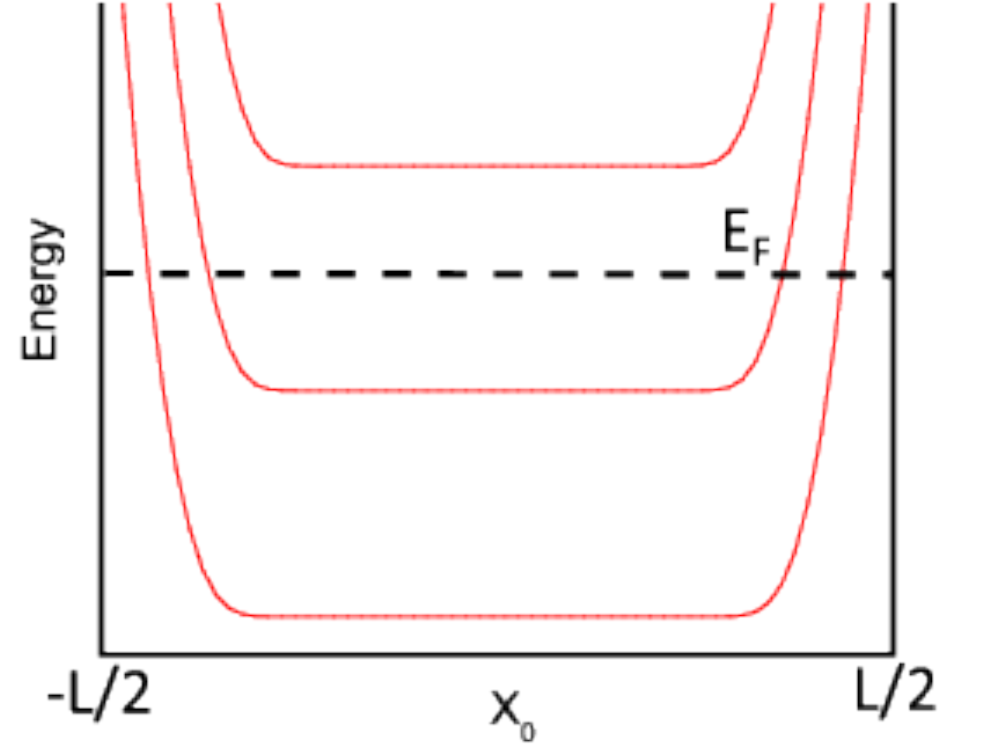}
\vskip-0.15cm
\caption{Sketch of the energy spectrum of the 2D electron gas in
perpendicular magnetic field with hard walls at $\pm L/2$.}
\end{figure}
Before addressing the problem of  open systems, we observe that the 
two above mentioned routes are not prepared to take into account
the role of the lead-sample coupling.
Still, Fig.1 offers a  hint, which usually is not remarked  
but is relevant in this respect: namely, the spatial distance of the  
edge states from the sample margin depends on their origin; 
more specifically, the states provided by the lower Landau bands are 
closer to the edges than those coming from the higher ones.
This suggests that the hybridization of the lead modes with the 
edge states is not unique, but depends on the gap where the edge states 
are energetically located. In this context, we shall show  how a poor 
hybridization affects  the  accuracy  of the conductance quantization. 

In the next section, we introduce the approach of the
effective non-Hermitian Hamiltonian for the open systems and show how it works for 
an analytically solvable toy model. Sec.III explains three transport regimes
of the 2DEG in strong perpendicular magnetic field in terms of the
lead-sample hybridization and  edge states lifetime, proving the  
presence of  superradiant effects at strong coupling, 
beyond the quantum Hall regime. The conclusions can be found in the last section.

\section{Effective Hamiltonian and one-dimensional toy model}
The Hamiltonian  of the composed system can be written as the sum 
of terms describing the 2D finite sample (S), the leads (L) and the 
sample-leads coupling:
\begin{equation}
H=H^S+H^L+\tau_c (H^{SL}+ H^{LS}), 
\end{equation}
where the  coupling parameter $\tau_c$ is  essential in the physics of 
open systems. From the point of view of the transport properties in 
strong magnetic field, the gradual increase of  $\tau_c$ moves the system 
from the resonant regime, expressed by sharp peaks of the transmission 
coefficient, to the quantum Hall (QH) regime described by  plateaus 
of the transmission (and, implicitly, of the Hall conductance). 
However, the study of some 1D open models indicates also the presence 
of a third regime in the limit of strong $\tau_c$, where superradiance 
effects occur \cite{Celardo, Zelevinski}. Such kind of effects should 
be expected  in the 2D open system  described by (1) as well, 
and one may ask what new spectral and transport features may  arise 
by increasing the coupling beyond the QH regime.

Superradiance phenomena in open systems are  described in terms of
non-Hermitian quantum mechanics, usually for one-dimensional models.
The main topic  of interest is  the evolution with the coupling 
strength of the complex eigenvalues, and of the  corresponding 
bi-orthogonal set of eigenfunctions. Effects as  resonance overlapping, 
segregation of the  eigenvalues or superradiance transition can be shown.
Specific superradiance effects are also revealed  by the quasiflat band in
phosphorene \cite{Bogdan}.

The problem described by (1) can be cast into a non-Hermtian formalism 
by including the effect of the leads into an effective Hamiltonian  
for the finite sample. In this formalism, the evolution of the eigenvalues
in the complex plane as function of $\tau_c$, combined with the degree 
of hybridization of the edge states with the lead modes, provides a 
unified description of the three transport regimes: resonant, 
quantum Hall and superradiant.
The approach allows to  visualize the  track of electrons
in the sample for all these regimes, and to get an understanding 
of the QHE beyond the heuristic picture. 

Technically speaking, the non-Hermitian effective Hamiltonian $H_{eff}^S$ 
that describes the finite sample in contact with leads can be obtained 
by the formal elimination of the leads degrees of freedom using a projection 
procedure. When projected on the lead and plaquette subspaces, 
the Hamiltonian (1) reads as the following $2 \times 2$ matrix:
\begin{equation}
\text{\textsl{\bf H}} =
\begin{pmatrix}
   H^{S}&\tau_cH^{SL}  \\
   \tau_cH^{LS}&H^{L}  \\
\end{pmatrix}.
\end{equation}

A way to prove the structure of $H_{eff}^S$ describing the plaquette 
in contact with the leads is via the Green function  defined as
${\bf G}(z)\big(z{\bf 1}-{\bf H}\big)={\bf 1}$ .
By projecting this equation on the two subspaces, the following system of 
equations is obtained:
\begin{eqnarray}
G^{SS}\big(z-H)_{SS}+G^{SL}(z-H)_{LS}=1 ,\\  \nonumber
G^{SS}\big(z-H)_{SL}+G^{SL}(z-H)_{LL}=0~ .
\end{eqnarray}
It results immediately 
$G^{SS}(z)\big(z-H^S_{eff}\big)=1$ with
\begin{equation}
H^S_{eff}(z)=H^S+\tau_c^2 H^{SL}\big(z-H^L\big)^{-1}H^{LS},
\end{equation}
where one notes that the effective Hamiltonian is non-Hermitian and 
dependent on the complex energy $z$. 

In order to continue, the explicit expressions of $H^S,H^L$ and 
$H^{LS}$ are necessary. In the tight-binding description, 
introducing the creation  (annihilation) operators $c_n^\dag (c_n)$ 
on the sites $\{n\}$ of the lattice, one may write:
\begin{equation}
H^S=\sum_{n,n'} t_{n,n'}e^{i\phi_{n,n'}}c^\dag_n c_{n'}, 
\end{equation}
where $t_{n,n'}$ is the hopping integral, and $\phi_{n,n'}$ stands 
for the Peierls phase in the presence of the  perpendicular magnetic 
field. The energy spectrum  of the confined (non-interacting, spinless) 
electronic system described by (5) is a Hofstadter-type butterfly,
where the energy gaps are filled with edge states.

While the  processes of interest occur in the plaquette, the role of the 
leads is to inject (and collect) independent electrons in the 2D plaquette. 
Along with B\"uttiker \cite{But}, each lead  contains several  independent 
channels, which are modeled by  semi-infinite one-dimensional chains:
\begin{equation}
\hskip-0.1cm
H^L=\sum_{\alpha=1}^4 H^L_{\alpha},~~ H^L_{\alpha}=t_L\sum_{\nu=1}^{N_c}
\sum_{i\ge 1}a^{\dag}_{\alpha,\nu,i}a_{\alpha,\nu,i+1} +H.c. ,
\end{equation}
where $\alpha$ counts the leads, $\nu$ counts the channels in the given 
lead, and $i$ stands for the site index in the tight-binding 
description of the chains. Next, the lead-sample coupling Hamiltonian 
can be written as:
\begin{equation}
H^{LS}=  \sum_{\alpha,\nu} c^{\dag}_{\alpha,\nu}a_{\alpha,\nu,1}~,
~~~H^{SL}=H^{LS\dag}. 
\end{equation}
This expression says that the first site $i=1$ of the channel $\nu$ in  the
lead $\alpha$ is sticked  to  the sample site denoted by ($\alpha,\nu$).
The summation over all channels and leads is performed.
The parameter $\tau_c$ represents the coupling strength and is  essential 
for the transport in meso-systems and superradiance physics in open systems.

Combining Eqs.(5-7) with Eq.(4), after some straightforward algebra,
the following explicit formula of  the effective Hamiltonian is obtained:
\begin{equation}
H^S_{eff}= H^S+\frac{\tau^2_c}{t_L} e^{ik}\sum_{\alpha,\nu}
c_{\alpha,\nu}^\dag c_{\alpha,\nu}~ .
\end{equation}
The above Hamiltonian, being non-Hermitian, exhibits complex eigenvalues. 
In addition, it depends on the energy $E$ of the incoming electrons via 
the quantity $k$ in (8) that parametrizes the energy as $E=2 t_L cos k$. 
\begin{figure}[htb]
\includegraphics[angle=-00,width=0.5\textwidth]{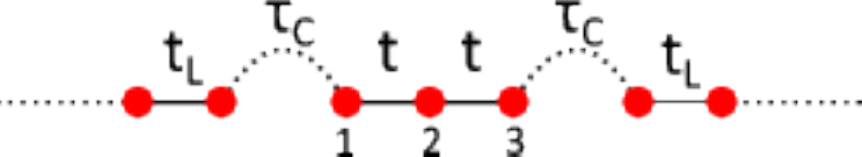}
\caption{Schematic description of a toy model with three sites.
The first and  last site are connected to semi-infinite leads. }
\end{figure}
At this stage, before addressing  the  problem of the 2D plaquette with 
several multi-channel leads, the discussion of a simple analytically 
solvable model is useful for understanding the influence of the 
lead-sample coupling on the complex spectrum and electron
transmission coefficient.
Let us consider  the  short chain composed of three sites, 
two of them being connected to semi-infinite leads via the parameter 
$\tau_c$, as in Fig.2. For this  system, the effective Hamiltonian  
reads in the matrix form as:
\begin{equation}
\text{\textsl{ $H^S_{eff}$}} =
\begin{pmatrix}
  \frac{\tau^2_c}{t_L}e^{ik} & t & 0  \\
   t & 0 & t  \\
   0 & t & \frac{\tau^2_c}{t_L}e^{ik} 
\end{pmatrix}.
\end{equation}

We have to remark that the non-Hermitian models mostly studied 
in the literature exhibit PT-symmetry \cite{Bender,Rotter}, 
in which case there are values of $\tau_c$ for which the spectrum 
is real. On the contrary, the effective Hamiltonian (9)  shows
broken PT-symmetry, and its spectrum is always complex, the
eigenvalues being:
\begin{equation}
\omega_0=i\tau^2_c/t_L,~~ 
\omega_{\pm}=\frac{1}{2}\big(\omega_0\pm \sqrt{8t^2-(\tau^2_c/t_L)^2)}~,
\end{equation}
where, for simplicity,  we have chosen $k=\pi/2$. 

The real and  imaginary part of the three eigenvalues are shown in Fig.3 
as function of $\tau_c$. One notices the coalescence
of the three branches of Re $\omega$, and also the  bifurcation
of Im $\omega$ at $\tau_c^2=2\sqrt{2} t t_L$, which is known  as 
an {\it  exceptional point} (EP) in the spectrum.
(We  noticed the EP occurs only for odd numbers of sites in the chain, 
but such aspects are beyond the main topic of the paper.)

It is important  to underline that the imaginary part of
two eigenvalues increases unboundedly with increasing $\tau_c$, 
generating what is called {\it superradiant} states. 
At the same time, the third one decreases to zero. 
In other words, the width $\Gamma= Im E$ of two resonances increases indefinitely, 
while the third one becomes thinner and thinner with increasing coupling. 
Such spectral properties are specific to superradiance, and our aim now 
is to  find out how they are reflected in the transport properties.
\begin{figure}[htb]
\hskip-0.5cm
\includegraphics[angle=-00,width=0.6\textwidth]{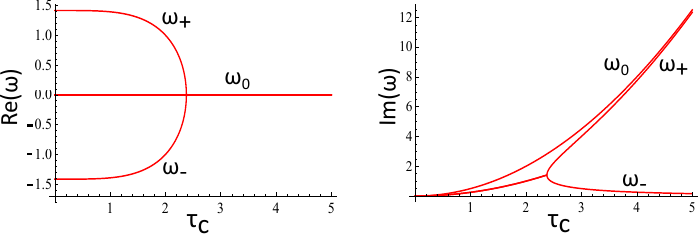}
\vskip-0.2cm
\caption{The real and imaginary part of the eigenvalues (10) as 
function of the coupling parameter $\tau_c$ ( $t=1$,  $t_L=2t$).}
\end{figure}
By calculating  the Green function that corresponds to the  Hamiltonian  (9),
the transmission coefficient is obtained immediately from
$T=4(\tau^4_c/\tau_L^2)|G^{SS}_{13}|^2 sin^2 k $, 
which is the  particularization of (11) for the case of only two contacts.
The result is depicted in Fig.4 as function of the gate potential and coupling 
parameter. (The Fermi level is fixed in the leads at $E_F=0$, 
and it is tuned to different spectrum domains by a gate potential  
applied on the sample, which can be varied continuously. The gate is  
simulated in the Hamiltonian by a diagonal term $V_g\sum_n  c^{\dag}_n c_n$.) 
\begin{figure}[htb]
\includegraphics[angle=-0,width=0.5\textwidth]{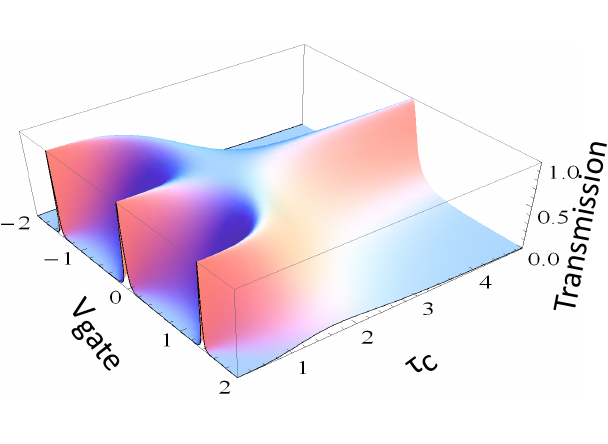}
\vskip-0.2cm
\caption{
The transmission coefficient $T$ of the toy model
as a function of the gate potential and coupling strength.}
\end{figure}
One has to observe in Fig.4 the presence of three resonant peaks at 
small $\tau_c$ and of a single narrow  peak that corresponds to the 
state $\omega_{-}$ at large $\tau_c$, showing that the  highly broadened
superradiant states do not contribute to the transmission. 
Another highly interesting thing is the formation of the plateau  
$T=1$ at a value of the  coupling which is below, but close to the 
exceptional point. Thus, we learn from this toy model
that the increase of the lead-sample coupling yields the broadening 
of the peaks, such that their overlapping give rise to a transmission plateau. 
Then, by increasing the coupling further on, the plateau is spoiled, 
and the system reenters a specific resonant regime, 
described by a single peak, where the  superradiant  states  are 
no more visible in the transmission.  

In what follows, we approach the problem of the  2D electron system in 
magnetic field, looking for similar effects and their implications for 
the quantum Hall effect.
\vskip-0.3cm
\begin{figure}[htb]
\includegraphics[angle=-90,width=0.5\textwidth]{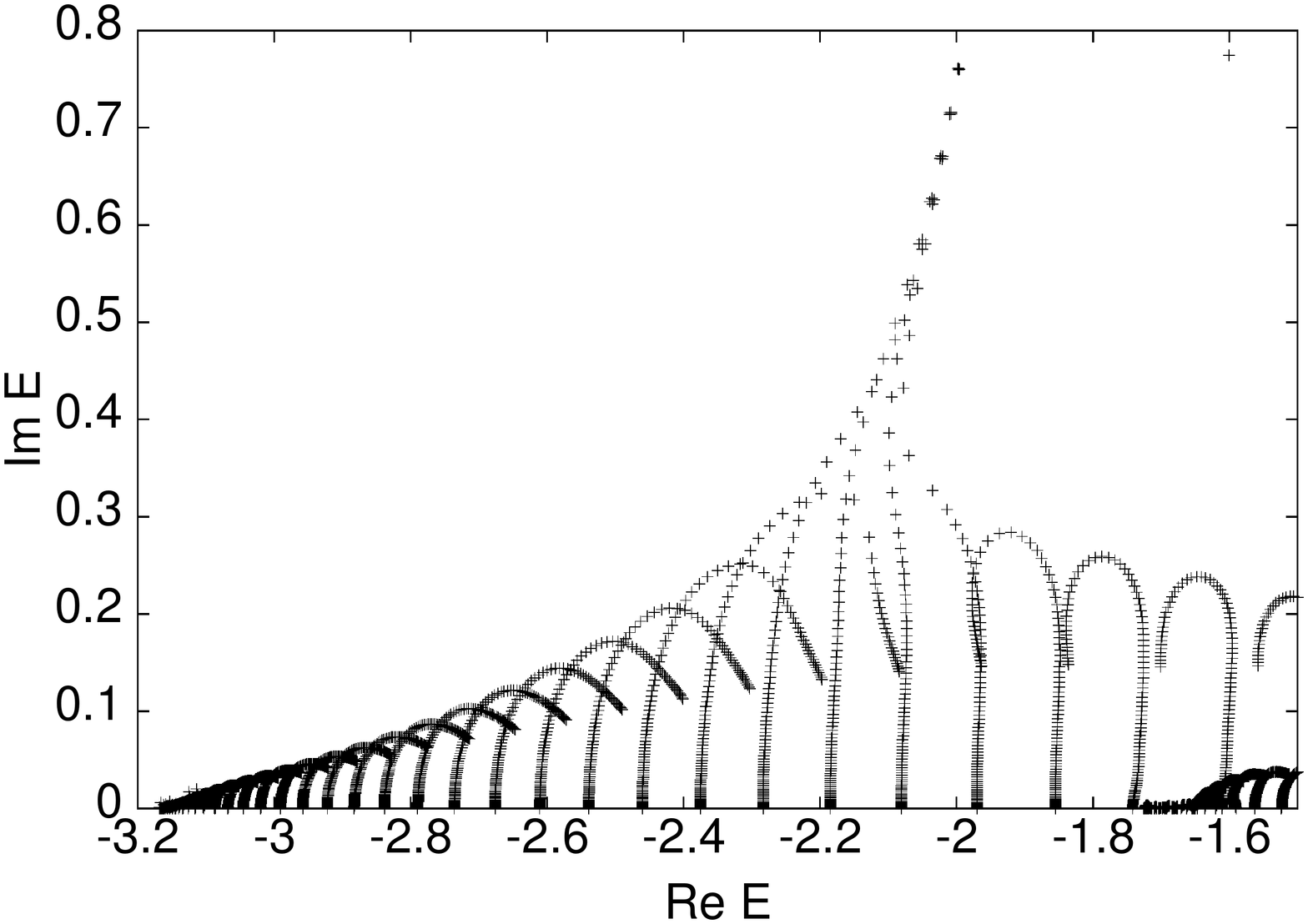}
\caption{
The evolution with increasing $\tau_c$ of the eigenvalues in the complex plane,
corresponding to edge states in the first gap. 
The turning point of the {\it hooks} occurs at $\tau_c\approx 1.5 t$. 
The superradiant states are also visible. (The plaquette dimension 
is 21$\times$21 sites, and the magnetic flux through the unit cell is 
$\Phi=0.15\Phi_0$, $\Phi_0$= magnetic flux unit.)}
\end{figure}

\section{Two-dimensional open system in strong magnetic field}
The first relevant effect  is the dependence on the coupling $\tau_c$
of the  complex eigenvalues of $H^S_{eff}$ shown in Fig.5.
It is to observe that, the leads being attached at the boarder 
of the plaquette, only the edge states eigenvalues are significantly 
affected. Both the real and imaginary part are  shifted with increasing 
$\tau_c$, two aspects being remarkable: i) the expected  initial
increase of $Im E$ is followed at stronger coupling
by a decrease that gives rise to a hook shape, and ii) the emergence of  
superradiant states, whose imaginary part increases unboundedly with 
increasing coupling. Both these  characteristics are  met also in 1D models
and considered  as typical for non-Hermitian Hamiltonians 
\cite{Zelevinski}.

The Hall conductance of a {\it four-lead} device can be derived 
using the Landauer-B\"uttiker formalism, which  involves the numerical 
calculation  of all transmission coefficients between different leads
$T_{\alpha,\alpha'}~(\alpha,\alpha'=1,..,4)$.
The formalism asserts that the formation of the IQHE plateaus is
conditioned by  integer values  of the transmission  coefficients 
$T_{\alpha+1,\alpha}$ between consecutive leads, and  zero 
values for all the others. It is obvious that  for pinched contacts 
(i.e., in the limit $\tau_c\rightarrow 0$) the system works in the 
resonant regime. The question then arises how strong the coupling should be  
in order for  the QH regime to occur.
In terms of the Green function, $T_{\alpha,\alpha'}$ is given by the 
known Caroli expression \cite{Caroli}:
\begin{equation}
T_{\alpha\alpha'}=\frac{4\tau_c^{4}}{t_L^2} \sum_{\nu,\nu'}
|G_{\alpha\nu,\alpha'\nu'}^{SS}|^{2} sin^2 k,~ \alpha\ne\alpha'.
\end{equation}
(Let us  remind that $sin k$ in (11) comes from the density of 
states in the leads.)
\begin{figure}[htb]
\includegraphics[angle=-90,width=0.5\textwidth]{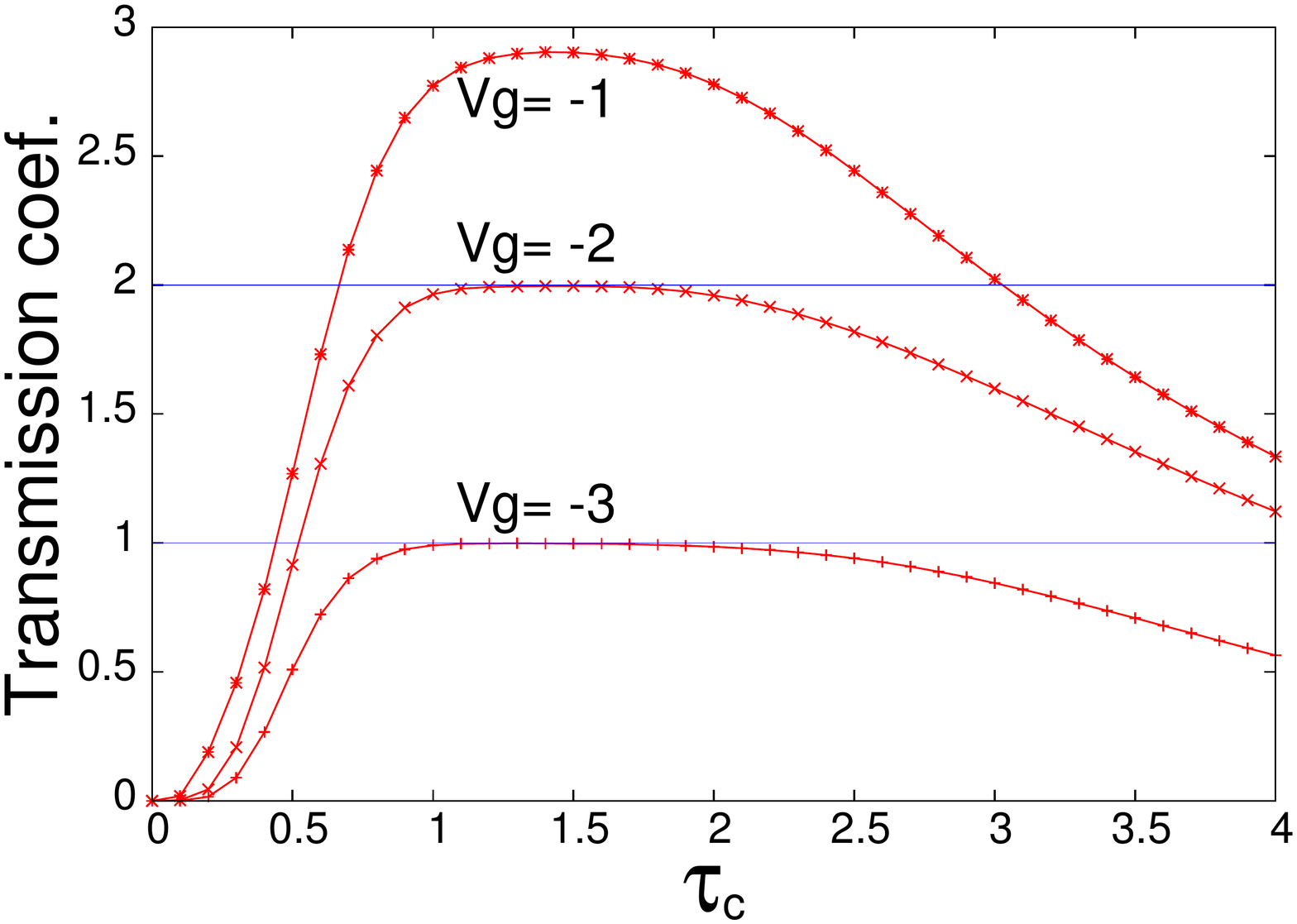}
\caption{$T_{21}$ as function of  $\tau_c$ for different values of 
the gate potential which fits the Fermi level in the first three gaps.
(The plaquette dimension is 40$\times$20 sites, $\Phi=0.1\Phi_0$.)}
\end{figure}
The result of the numerical calculation of $T_{21}$ as function 
of $\tau_c$ is shown in Fig.6 for three different positions of
$E_F$ in the first three gaps.
One should notice that  the range of $\tau_c$ that ensures 
the transmission quantization shrinks when moving to higher gaps, 
and already in the third one the expected value $T_{21}=3$ is no more 
perfectly reached.
This effect is symptomatic, and it can be understood in terms of
imperfect channels due to insufficient hybridization of the lead modes
to the innermost edge state.The smaller the plaquette is, the easier 
this outcome can be evidenced.
One has also to observe the decrease of the transmission 
at high coupling, announcing the decay of the IQHE and 
the reentrance of a resonant-type regime. 
Aiming to understand the result in Fig.6, we  discuss the life time 
defined by the inverse of the energy imaginary part $\tau=\hbar/Im E$, 
and also the path followed by the electron inside the sample for 
the different transport regimes we come across.

Let us assume that the electron enters the sample
along a given lead $\alpha$. The coupling term $H^{SL}+H^{LS}$ hybridizes
the lead eigenfunction $|\Phi^L_{\alpha}\rangle$ with those of the sample
and generates the scattering wave function $|\Psi_{\alpha}\rangle$
described by the Lipmann-Schwinger expression:~~
$|\Psi_{\alpha}\rangle = |\Phi^L_{\alpha}\rangle + GV|\Phi^L_{\alpha}\rangle$,
with $V=H-H^L_{\alpha}$. Then, the electron path through the sample can be 
visualized by the projection of $|\Psi_{\alpha}\rangle$ on {\it all} sample sites  
$|\langle n|\Psi_{\alpha}\rangle|^2$. As technical aspects, we mention that 
from all terms in $V$ only $H^{SL}$ contributes, while the Green function $G$, 
being projected on the sample sites, should be replaced by $G^{SS}$. 
One obtains readily: 
\begin{equation}
|\langle n|\Psi_{\alpha}\rangle|^2= \tau_c^2\sum_{\nu}| \langle n|
\frac{1}{V_g-H^S_{eff}}|\alpha,\nu\rangle|^2,~~  n\in S,
\end{equation}
where $|\alpha,\nu\rangle=c^{\dag}_{\alpha,\nu} |0\rangle$, and again
$V_g$  is the gate potential.

Let us discuss first the resonant regime. In this case, the edge states of 
the isolated sample (described by $H^S$) are weakly perturbed by the leads, 
so that the imaginary part acquired by the complex eigenvalues of $H^S_{eff}$
is small 
(more precisely, $Im E << \Delta$, where $\Delta$ is the mean 
level spacing  between consecutive edge states). 
Correspondingly, the life time of the injected electron in the 
sample is long, meaning that the electron can get out of the sample
with significant probability at any contact. 
The consequence is that, in the resonant regime, all $T_{\alpha,\alpha'}$ 
are different from zero. Further support to this picture comes from  
the charge distribution on the plaquette given by Eq.(12), 
which turns out to be distributed  all around the perimeter 
(as shown in Fig.7a), similarly to the edge state in the isolated plaquette. 
(The life time is the time spent by electron in the sample, so that
equivalently one may use also the term 'escape time' as in \cite{Datta}.)

The increase of $\tau_c$ results in reducing the life time 
and shortening 
the electron path in the sample as shown in Fig.7b.
Expecting to reach the QH regime, we  increase  the coupling further on, and, 
indeed, one arrives at the situation  when the scattering wave that enters 
the contact $\alpha$ leaves the sample at the next contact $\alpha+1$, 
as obvious in Fig.7c. Then, $T_{\alpha+1,\alpha}$ is the only non-vanishing 
transmission coefficient and equals an integer.
This is just the requirement for the realization of the quantum plateau. 

The construction of the scattering wave function is different for the two cases 
discussed above. If $ImE << \Delta$, the density of energy states in the sample
shows a peak structure, and the scattering wave function results from the 
hybridization of the lead mode with the  edge state that satisfies the 
resonance condition. On the other hand, if  $ImE \sim \Delta$, 
due to the level broadening, the scattering wave function  is built by  
weighted contribution of different edge states.
In this case  (which  occurs by increasing  the coupling $\tau_c$),
the density of states  becomes a smooth function of energy in the whole 
energy range covered by edge states, and the same occurs for the transmission
coefficient, meaning that we have reached the  QH regime. As we already mentioned, 
the toy model reveals a similar  behavior of the transmission in Fig.4.
\begin{figure}[htb]
\includegraphics[angle=-00,width=0.7\textwidth]{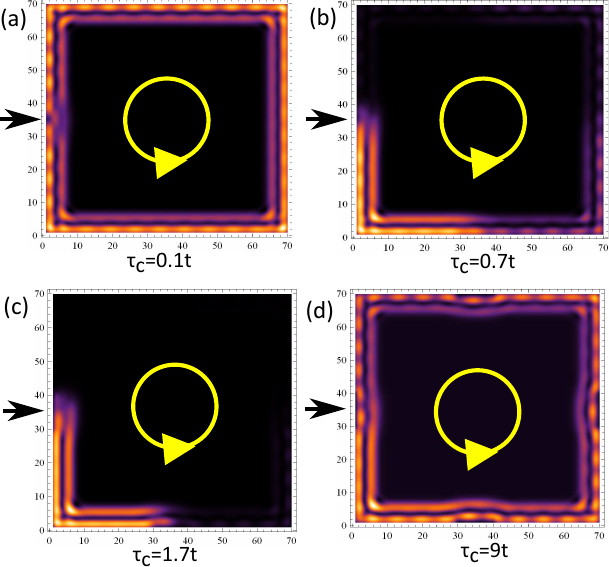}
\caption{The path (in red) followed by electrons in 
a square plaquette with four leads for different couplings  $\tau_c$, 
which correspond to different regimes. The injection place
is indicated by  black, and the path chirality by  yellow arrows.
$E_F$ is tuned in the second gap. (The plaquette dimension is 70$\times$70 
sites, each lead contains 10 channels, and $\Phi=0.04\Phi_0$.)}
\end{figure}

The  limit of the very strong coupling, known in the physics of 
open systems as the superradiant regime, was not yet studied  from 
the point of view of the quantum transport properties in 2D systems. 
We know already from Fig.5 and Fig.6 that, for high values of $\tau_c$,
the life time increases again, while the transmission coefficient 
decreases below the integer values.
In addition, one may observe in Fig.7d that the electron path
is  enlarged again and keeps the plaquette perimeter 
similarly to the resonant regime. These circumstances suppress the IQHE 
plateaus, and the numerical calculation  indicates that they are 
replaced by transmission oscillations when $V_g$ is varied.
\begin{figure}[htb]
\includegraphics[angle=-00,width=0.45\textwidth]{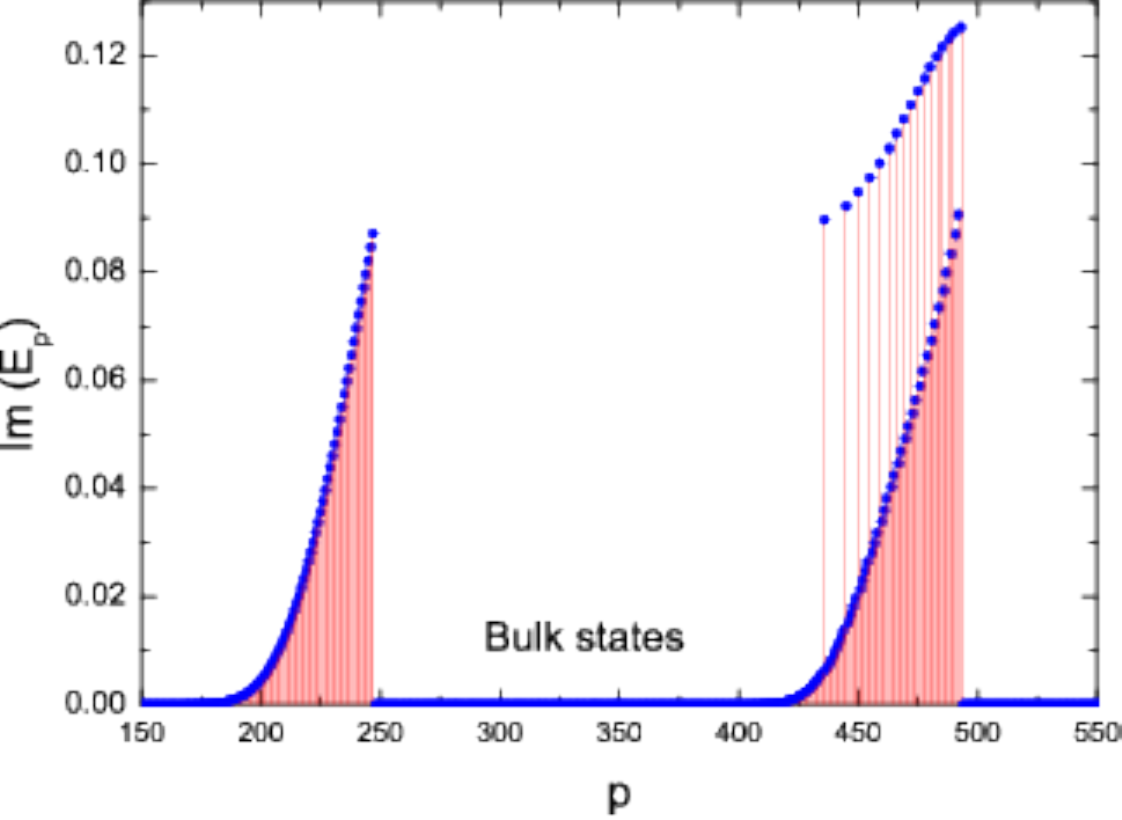}
\vskip-0.2cm
\caption{The imaginary part $ImE_p$ as function of the eigenvalue
index $p$ in the spectral range of the first two gaps. 
The parameters are the same as in Fig.7,  $\tau_c=1.7 t$.}
\end{figure}

We are still left with the question concerning the formation of
many channels in the higher gaps; without lost of generality,
we restrict the discussion to the second one. Fig.8 strings the widths 
of all edge states in the first two gaps. The apparently empty range 
$p\in[250,420]$ is in fact occupied by bulk band  states which, 
as mentioned already, show vanishing width. 
The important issue is that the values of $Im E_p$ in the second gap 
segregate into two sets, proving the existence of two types 
of edge states distinguished by their width.
The calculation being performed at given $\tau_c$, we conjecture that 
the segregation originates in the different hybridization
of different types of edge states existing in the gap.
Obviously, the  higher widths comes from a stronger hybridization
and should correspond to edge states localized closer to the margin. 
According to the hint given by the heuristic model Fig.1,
they can  be associated to the edge states  detached from the
first Landau band. 
Similar reasoning says that the family of states with
low width are more remote from the margin and correspond to the 
states provided by the second band.

One may further speculate that in the higher gaps there are edge states 
localized far away from the margin, having a  small overlap with the lead 
modes and a  long life time. 
Then, the corresponding channels do not get out of the sample 
at the next contact and spoils the quantized value of the transmission, 
giving rise to imperfect QHE plateaus of higher index. 
This is the case in Fig.6 for the third gap. 
The appearance of such deviations from the very well  quantized plateaus
depends on different parameters like the plaquette dimension, 
value of $\tau_c$ or  magnetic field strength.

\section{Conclusions}
In summary, we highlighted the determinant role of the  edge states 
lifetime in defining different transport regimes in open 
systems: resonant, quantum Hall and superradiant.
The lifetime is defined by the imaginary part of the complex eigenvalues 
of the non-Hermitian effective Hamiltonian  that describes the 2D finite 
sample in contact with leads. The lead-sample coupling induces  a specific 
evolution of the eigenvalues in the complex plane and controls 
the conditions for emergence of the IQHE and superradiant regime 
(which appears {\it beyond} the QH regime, in the limit of high coupling).
We used the model Hamiltonian (1), however we think that the 
qualitative physical results are generally valid.

Although deduced for finite 2D electron gas in magnetic field, 
our considerations are applicable to all  materials exhibiting
edge/surface states, including  those recently discovered in the 
topological insulators.

A difficult question is under what circumstances the superradiant 
regime can be seen experimentally. In principle, the solution is to  
decrease the hopping parameter $t$, that is to use narrow band materials.
However, at that moment the correlations become important, 
and the one-particle description used here  becomes questionable.
Of course, the superradiant effects combined with the fractional Hall effect
would be highly interesting, however it is much beyond our present aim.
Then, an alternative would be to decrease $t_L$, meaning leads with heavy 
electron effective mass.
\section{Acknowledgments}
We acknowledge the financial support from Romanian Core Research 
Programme PN16-480101.

\end{document}